\documentclass[prb,showpacs,preprint]{revtex4}
\usepackage{epsf}
\begin{document}

\title{Magnetoresistance behavior of a ferromagnetic shape memory alloy: Ni$_{1.75}$Mn$_{1.25}$Ga} 
\author{S. Banik$^1$, R. Rawat$^1$, P. K. Mukhopadhyay$^2$,  B. L. Ahuja$^3$, Aparna Chakrabarti$^4$, P. L. Paulose$^5$, S. Singh$^1$, A. K. Singh$^6$, D. Pandey$^6$, and S. R. Barman$^{1*}$}
\affiliation{$^1$UGC-DAE Consortium for Scientific Research, Khandwa Road, Indore, 452017, India}
\affiliation{$^2$LCMP, S. N. Bose National Centre for Basic Sciences, Kolkata, 700098, India}
\affiliation{$^3$Department of Physics, M. L. Sukhadia University, Udaipur 313001, India}
\affiliation{$^4$Raja Ramanna Centre for Advanced Technology, Indore, 452013, India}
\affiliation{$^5$Tata Institute of Fundamental Research, Homi Bhabha Road, Mumbai, 400005, India.}
\affiliation{$^6$School of Materials Science and Technology, Banaras Hindu University, Varanasi, 221005, India.} 

\begin{abstract}  
A negative$-$positive$-$negative switching behavior of magnetoresistance (MR) with temperature is observed in a ferromagnetic shape memory alloy Ni$_{1.75}$Mn$_{1.25}$Ga.
~In the austenitic phase between  300 and 120\,K,  MR is negative due to $s-d$ scattering. Curiously, below 120\,K   MR  is positive, 
~ while  at still lower temperatures in the martensitic phase, MR is  negative again.
~  The positive MR cannot be explained by Lorentz contribution and is related to 
 ~a magnetic transition. Evidence for this is obtained from {\it ab initio} density functional theory, a decrease in magnetization and resistivity upturn  at 120~K. 
 ~ Theory shows that a ferrimagnetic state with anti-ferromagnetic alignment between the local magnetic moments of the Mn atoms 
 ~is the  energetically favoured ground state.  In the martensitic phase, there are two competing factors that govern the MR behavior: a dominant negative trend up to the saturation field due to the decrease of electron scattering at twin and domain boundaries;  and a weaker positive trend due to the ferrimagnetic nature of the magnetic state. MR exhibits a hysteresis between heating and cooling that is related to the first order nature of the martensitic phase transition.
\end{abstract}
 
\maketitle

\section{Introduction}
 Recent years have witnessed extensive research on magnetoresistance (MR) to understand its basic physics in metallic multilayers,  transition metal oxides, {\it etc}.\cite{Baibich88} Ferromagnetic shape memory alloys (SMA) are of current interest because of their  potential technological  applications and the rich physics they exhibit.\cite{Kainuma06,Krenke05,Takeuchi03,Planes97,Boon07,Chernenko04,Barman05,Chakrabarti05,Barman07,Ahuja07,Banik06} Large magnetic field induced strain (MFIS) of 10\% with actuation that is faster than conventional SMA's has been obtained in Ni-Mn-Ga.\cite{Murray00,Sozinov02} MFIS is achieved by twin boundary rearrangement in the martensitic phase and the main driving force for twin boundary motion in the presence of a magnetic field is the large magnetocrystalline 
anisotropy (MCA).\cite{Murray00,Sozinov02,Albertini01,Ullakko01,Pan00,Sullivan04,Sozinov04} 

Negative MR has been observed earlier in SMA's like Cu-Mn-Al and was associated with the possible presence of
Mn-rich clusters in the Cu$_2$AlMn structure.\cite{Marcos02} Recently, we have reported a negative MR of about 7.3\% at 8~T at room temperature in Ni$_{2+x}$Mn$_{1-x}$Ga.\cite{Biswas05} It was explained by $s-d$ scattering model for a ferromagnet, while the differences in the MR behavior in the martensitic phase compared to the austenitic phase was related to twin variant rearrangement with magnetic field.\cite{Biswas05} MR ranging between -1 to -4.5\% has been reported for thin films of Ni-Mn-Ga.\cite{Lund02,Golub04} Recently, a large negative MR of 60-70\% has been reported for Ni-Mn-In, which has been explained by the shift of the martensitic transition temperature with magnetic field.\cite{Yu06,Sharma06}  

Ni$_{2-y}$Mn$_{1+y}$Ga with $y$=\,0.25 {\it i.e.} Ni$_{1.75}$Mn$_{1.25}$Ga is one of the unique compositions in the Ni-Mn-Ga family that has low martensitic transition temperature ($M_s$) of about 76\,K.\cite{Liu05} This enables the study of the ferroelastic transition  much below the Curie temperature ($T_C$=\,380~K). 
Here, part of the Mn atoms ($y$=\,0.25), referred to as MnI, occupy the Ni site while the remaining Mn ($y$=\,1.0) atoms at the Mn site are referred to MnII. The MnI atoms, which are 20\% of the total Mn atoms, are excess with respect to the stoichiometric Ni$_2$MnGa composition. These excess MnI type atoms are expected to have interesting influence on the  resistivity, MR, and magnetization, since in related systems like Ni$_2$Mn$_{1.25}$Ga$_{0.75}$ and Mn$_2$NiGa their moments are reported to be anti-parallel to the MnII atoms.\cite{Barman07,Enkovaara03}  
~Here, we report  an intriguing switching behavior of MR with temperature that is related to the occurrence of martensitic transition at low temperature in the ferrimagnetic state. To the best of our knowledge, such MR behavior reported here has not been observed in any magnetic material till date. This basically arises from the interplay of magnetism and shape memory effect. Our studies indicate possibility of  new practical applications for  ferromagnetic SMA  as  magnetic sensor for data storage and encryption, whose response can be toggled by changing the  temperature. It is envisaged that the multifunctional combination of properties (magnetic  sensing, magnetocaloric, actuation and shape memory effects) of the ferromagnetic SMA's will be important for their future application.

\section{Methods}
Bulk polycrystalline ingots of  Ni$_{1.75}$Mn$_{1.25}$Ga have been prepared by the standard method of melting appropriate quantities of Ni, Mn and Ga (99.99\% purity) in an arc furnace. The ingot was annealed at 1100\,K for nine days for homogeneization and subsequently quenched in ice water.\cite{Banik06,Banik07a} The composition has been determined by energy dispersive analysis of x-rays using a JEOL JSM 5600 electron microscope. 
~A superconducting magnet from Oxford Instruments Inc., U.K. was used for carrying out the longitudinal MR measurements up to a  maximum magnetic field of 8~T.\cite{Rajeev}  MR  is defined as $\Delta\rho_{m}$=~$\frac{\Delta\rho}{\rho_0} = \frac{(\rho_H-\rho_0)}{\rho_0}$,   where $\rho_H$ and $\rho_0$ are the resistivities in  $H$ and zero field, respectively. The statistical scatter of the resistivity data  is 0.03\%. $M(T)$ measurements were performed with Lakeshore 7404 vibrating sample magnetometer with a close cycle refrigerator. $M(H)$ measurements were done using a MPMS XL5 SQUID magnetometer. Temperature-dependent powder x-ray diffraction (XRD) data were collected using an 18 kW copper rotating anode-based Rigaku powder diffractometer fitted with a graphite monochromator in the diffracted beam. The temperature was stable within $\pm$0.3~K during data collection at each temperature. The data were collected in the Bragg- Brentano geometry using a scintillation counter.

Spin-polarized first principle density functional theory  calculations were performed by full potential linearized augmented plane wave (FPLAPW) method using the WIEN97 code.\cite{Wien97} Generalized gradient approximation for the exchange correlation was used.\cite{Perdew96} The muffin-tin radii were taken to be Ni: 2.1364~a.u., Mn: 2.2799~a.u., and  Ga: 2.1364~a.u. The convergence criterion for total energy was  0.1 mRy, {\it i.e.} an accuracy of $\pm$0.34 meV/atom. The details of the  method of calculation are given elsewhere.\cite{Barman07,Chakrabarti05}

\section{Results and Discussion}


Fig.~1 shows the isothermal magnetoresistance ($\Delta\rho_{m}$) of  Ni$_{1.75}$Mn$_{1.25}$Ga as a function of magnetic field at different temperatures. It can been seen from the figure that at 300~K, the magnitude of $\Delta\rho_{m}(H)$ increases with $H$ to -1.35\% at 8\,T (Fig.\,1a). In order to ascertain the H dependence, we have fitted $\Delta\rho_{m}(H)$ by a second order polynomial of the form $\alpha$\,H\,+\,$\beta$\,H$^2$(solid lines in Fig.~1). We find  the second order term ($\beta$) to be very small, the ratio $\beta/\alpha$ being 0.02, which shows that the variation is essentially linear. Similar linear variation is obtained up to 150\,K, although the magnitude of $\Delta\rho_{m}$ decreases to -0.3\%.   Linear variation of negative MR with field has  been observed for Ni$_2$MnGa.\cite{Biswas05} Also Kataoka has calculated $\Delta\rho_{m}(H)$ for ferromagnets with different electron concentrations using the $s-d$ scattering model, where the scattering of $s$ conduction electrons by localized $d$ spins is suppressed by the magnetic field resulting in a decrease in $\rho$.\cite{Kataoka01} Magnitude of $\Delta\rho_{m}$ is shown to increase almost linearly with $H$ for  ferromagnetic materials.\cite{Kataoka01} Since Ni$_{1.75}$Mn$_{1.25}$Ga has large Mn 3$d$ local moment with high electron concentration (valence electron to atom ratio, $e/a$=\,7.31), negative MR in the 150-300~K range is well described by the $s-d$ scattering model. As the temperature is lowered, $\Delta\rho_{m}(H)$ decreases due to reduction in the spin disorder scattering. 
  

MR in Fig.~1b shows an interesting behavior:  $\Delta\rho_{m}(H)$ is positive at 100~K. However, at 50~K it is negative, but with a different $H$ dependence compared to the $s-d$ scattering regime (Fig.~1a). In other ferromagnetic Heusler alloys like Ni$_2$MnSn and Pd$_2$MnSn, positive MR has been observed and attributed to the Lorentz contribution.\cite{Hurd81}  In such cases, $\Delta\rho_{m}$ is positive at lowest temperatures and decreases as temperature increases. For example, MR is positive for Pd$_2$MnSn  at 1.8\,K and is negative above 60\,K.\cite{Hurd81} In contrast, the MR variation in  Ni$_{1.75}$Mn$_{1.25}$Ga is opposite. Lorentz contribution gives rise to a positive MR when the condition $\omega{_C}\tau>>1$ is satisfied, where $\omega{_C}$ and $\tau$ are cyclotron frequency and conduction electron relaxation time, respectively. This condition is valid for extremely pure metallic single crystals at very low temperatures (where $\tau$ is large and $\rho$$\leq$10$^{-8}\,\Omega cm$) or at large $H$ (where $\omega{_C}$ is large). But for Ni$_{1.75}$Mn$_{1.25}$Ga, the residual resistivity is large, implying small $\tau$ so that even at 8\,T   the above condition is not satisfied. By the same argument, we expect a more positive contribution at 5\,K compared to 50\,K, since the resistivity is lower at 5\,K (Fig.~2a). On the other hand, the observed data show opposite trend. Hence, the positive MR in Ni$_{1.75}$Mn$_{1.25}$Ga cannot be ascribed to Lorentz force and  other mechanisms need to be explored to understand this finding.

Fig.~2a shows resistivity ($\rho$$(T)$) at zero and 5~T magnetic field between 5 and 180\,K for two cycles. 
Above 120\,K, where the sample is in the austenitic phase $\rho$$(T)$ has a positive temperature coefficient of resistance, and the data for the different cycles overlap. Between 88 and 37\,K, the hysteresis in $\rho(T)$ becomes highly pronounced and this is a signature of the  martensitic transition. The martensitic transition is also clearly shown by the ac-susceptibility data in Ref.\onlinecite{Liu05} and the low field magnetization data shown in Fig.~4a (discussed latter). 
~The onset of the martensitic transition is depicted by the change in slope in $\rho$ at $M_s$ (=\,76\,K, in agreement with Ref.\onlinecite{Liu05}). The other transition temperatures like  martensite finish $M_f$=\,37\,K, austenitic start $A_s$=\,47\,K, and austenitic finish $A_f$=\,88\,K, shown in Fig.~2a, concur with the $M(T)$ data to be discussed later (Fig.~4a).   $\rho$ shows a step centered around 65\,K. This possibly arises due to strain effect on the  nucleation and growth of the martensitic phase at such low temperatures, and similar effect has been observed in Ni$_2$FeGa.\cite{Liu05}


In order to establish beyond any doubt that the hysteresis in $\rho$$(T)$ is related to the martensitic transition, we show the powder XRD pattern at different temperatures in Fig.~3. To record the XRD patterns, Ni$_{1.75}$Mn$_{1.25}$Ga ingot was crushed to powder and annealed at 773~K for 10 hrs to remove the residual stress. The $L_{2_1}$ cubic austenitic phase is observed up to 100~K.  There is no signature of any phase transition, related to the formation of  a possible premartensitic phase around 120~K, which could have been responsible for the upturn in $\rho$$(T)$. 
~The lattice constant at 100~K turns out to be $a_{aus}$=\,5.83\AA.~ At 80~K, new peaks appear. These peaks correspond to the martensitic phase and coexist with the austenite peaks. By 40~K, the XRD pattern shows that the martensitic transition is  complete as there is no austenite phase, in agreement with the $\rho(T)$ data.
~The XRD patterns have been indexed by Le Bail fitting procedure;\cite{Lebail88} and we find that the martensitic phase is  monoclinic in the $P2/m$ space group. The refined lattice constants are $a$=\,4.22, $b$=\,5.50 and $c$=\,29.18\,\AA,~ and $\beta$=\,91.13. Since $c$$\approx$\,7$\times$$a$, a seven layer modulation may be expected, and such modulated structures with monoclinic or orthorhombic symmetry have been reported for  Ni-Mn-Ga.\cite{Brown02}  Magnetic field induced strain has been observed in Ni-Mn-Ga for structures that exhibit modulation.\cite{Murray00,Sozinov02} The unit cell volume of the martensitic phase is within 2\%  of that of the equivalent austenitic cell given by 7$\times$$a_{aus}^3$/2. This shows that the unit cell volume  changes little between the two phases, as expected for a shape memory alloy.\cite{Bhattacharya03} 


After establishing the existence of the structural martensitic transition from XRD, we discuss the details of the resistivity behavior. $\rho(T)$ at 5T shows a difference in the first and second FH (field heating) cycles, the first cycle $\rho(T)$ being higher. In the first FH cycle, the sample is subjected to a magnetic field of 5T at 5~K after ZFC (zero field cooling).  Subsequently, FC data were taken and then the second cycle of FH was measured. Thus, while in FH first cycle, the magnetic field of 5~T was switched on at 5K, where as in the FH second cycle the field is on from RT.  The possible reason for the difference in resistivity between the two FH cycles is discussed later on. 
In Fig. 2b, we show the MR calculated from the difference between the ZFC and FC (cooling MR data) and ZFH and FH second cycle (heating MR data). But, if the FH first cycle is considered, MR is lower by about 0.4\% at 5~K, which  argees with the value in Fig.~1. This is because the MR in Fig. 1, $\Delta\rho_{m}$ is measured in a different way: at 300~K, H is varied from 0 to 8 to 0 to -8 to 0 Tesla. Then, the specimen was cooled down to the next measurement temperature of 235~K and the field was varied in a similar way. For the next measurement, the sample was heated up to 300~K and cooled under zero field condition to the  temperature of measurement. Thus, for the martensitic phase, this MR data (Fig.~1) can be related to $\Delta\rho_{m}$ calculated from ZFH and FH first cycle $\rho$ data (Fig.~2a). 

Fig.~2b clearly shows the switching effect in MR as a function of temperature. A comparison of Fig.~2a and b shows a significant correlation between the hysteresis in MR and $\rho$$(T)$. For the cooling cycle, MR is negative from 300 to 135~K, and exhibits a negative to positive switching at 135~K. This negative MR region is explained by $s\,-\,d$ scattering, as discussed above. MR is positive between 135 to 76~K (=\,$M_s$); and this is also manifested in MR(H) data at 100~K in Fig.~1b. As discussed earlier, the positive MR cannot be assigned to Lorentz force contribution.
MR exhibits a positive to negative switching at $M_s$ in the cooling cycle.  MR becomes negative at M$_s$ with a shallow minimum at 73~K,
~shows a hump at 64~K,
~and then plunges to large negative values below 64~K, and finally increases slightly to reach a temperature independent value of about -3\% below 37~K~(=\,$M_f$). The shape of the MR(T) curve in Fig.~2b during heating is very similar to that during cooling, but is shifted in temperature in the martensitic transition region due to hysteresis.
Thus, hysteresis in MR is clearly observed, which 
~indicates the possibility of studying phase co-existence and first order phase transition in FSMA's using MR.





Magnetization measurements have been performed to understand the magnetoresistance behavior. A sharp decrease of magnetization at the martensitic transition (Fig.~4a)  in small fields (0.01-0.1~T) is the manifestation of  large MCA  in the  martensitic phase.\cite{Albertini02} Large MCA has been observed in different Ni-Mn-Ga alloys and is responsible for magnetic field induced twin variant reorientation. The magnetization in the martensitic phase decreases because 
~in the low field,  a twinned state with moments along the  easy axis ([001]) oriented in dissimilar directions for different twins is energetically favorable. The gradual decrease of magnetization in the austenitic phase, on the other hand,  is possibly related to an increase of the austenitic phase MCA with decreasing temperature.\cite{Albertini02} A step-like decrease in magnetization with distinct change of slope is evident at 120\,K  for both 0.01 and 0.1 T fields (inset, Fig.~4a).
~This decrease is significant because it  suggests that the upturn in $\rho$(T) and positive MR could be related to a  magnetic transition that decreases the magnetization.

Fig.~4b shows $M(T)$ at 5~T in FC and FH. This field is much higher than the saturation field, as shown by the isothermal $M-H$ curves in Fig.~5. $M(T)$ shows the characteristic variation of saturation magnetization with temperature. By fitting the higher temperature region using the expression ($T_C$\,-\,T)$^\gamma$ (bold dashed line in Fig.~4b), we obtain an approximate estimate of $T_C$ to be 380~K.
~This is close to the $T_C$ of 385\,K reported for Ni$_{1.8}$Mn$_{1.2}$Ga.\cite{Liu05} In comparison to Fig.~4a, magnetization increases by two orders of magnitude for 5~T FC and FH runs. Thus, the changes in the magnetization that are clearly visible in the low field measurement (Fig.~4a) are not evident here. 
~For example,  the large relative decrease in magnetization in the martensitic phase (Fig. 4a) and the decrease at 120~K are not observed in Fig. 4b. 
~Instead, the magnetization gradually increases in the martensitic phase. This increase is intrinsic and is due to higher saturation magnetization in the martensitic phase. This results from alterations in interatomic bonding related to the change of structure, as also observed in Ni$_2$MnGa.\cite{Barman05,Webster84} The saturation moment turns out to be 3.5~$\mu_B$. The only signature of the martensitic transition in Fig.~4b is the hysteresis in $M(T)$ during heating and cooling cycles. 
~However, there is hardly any change in the martensitic transition temperature with field. This shows that for this alloy, magnetic field does not change $M_s$ resulting in magnetic field induced martensitic transition, unlike in Ni-Mn-Sn and Ni-Mn-In.\cite{Krenke05} 
~The isothermal $M-H$ loop in Fig.~5 shows a decrease in the saturation magnetic field between the martensitic phase (20~K) and the austenitic phase (283 or 360~K). This is because in the austenitic phase, the MCA is very small and there is no twinning compared to the martensitic phase with large MCA and twinning.

We have calculated the magnetic ground state using {\it ab initio}, spin polarized density functional theory employing FPLAPW method to understand the origin of positive  MR behavior. Good agreement between  experiment and  theory has been obtained earlier for the magnetic moments, lattice constants, total energies and the density of states for both the phases.\cite{Chakrabarti05,Banik06,Banik07a,Barman05,Barman07} In particular, the total energies have been used to explain the phase diagram and magnetic states of Ni$_2$MnGa, Ni$_{2.25}$Mn$_{1.75}$Ga and Mn$_2$NiGa.\cite{Chakrabarti05,Barman07}

Here, we calculate the total energies of the different magnetic states of non-stoichiometric Ni$_{1.75}$Mn$_{1.25}$Ga for the $L_{2_1}$ structure (see Fig.~1 in Ref.\onlinecite{Barman05}) with lattice constant of 5.843\AA~ determined from XRD at room temperature. The structure consists of 4 interpenetrating f.c.c. sub-lattices occupied by two Ni atoms, one Mn (MnII) and one Ga atom. To emulate the non-stoichiometric composition, a 16 atom $L_{2_1}$ super-cell is considered, where one of the eight Ni atoms is replaced by one excess MnI type atom.\cite{Chakrabarti05} Thus,  there are seven Ni, five Mn (one MnI and four MnII {\it i.e.} out of total Mn atoms only 20\% are MnI) and four Ga atoms in the super cell with the chemical formula  Ni$_7$Mn$_5$Ga$_4$, which is equivalent to Ni$_{1.75}$Mn$_{1.25}$Ga.  The total energy that consists of the total kinetic, potential, and exchange correlation energies of a periodic solid with frozen nuclei has been calculated for two magnetic configurations with MnI spin moment parallel and anti-parallel to MnII. We find that the total energy is significantly lower by 16 meV/atom, 
~when MnI is anti-parallel to MnII, compared to their parallel orientation. Anti-parallel alignment of Mn spins is energetically favored because of the direct MnI-MnII nearest neighbor (at 2.53\,\AA~ distance) interaction, as has been shown for other Mn excess systems like Mn$_2$NiGa, Ni$_2$Mn$_{1.25}$Ga$_{0.75}$ and Ni-Mn-Sn.\cite{Krenke05,Barman07,Enkovaara03} The exchange pair interaction as a function of Mn\,-\,Mn separation  was calculated by a Heisenberg-like model and an antiferromagnetic coupling at short interatomic distances was  found.\cite{Hobbs03}  Enkovaara {\it et al.} reported antiferromagnetic Mn configuration in  Ni$_2$Mn$_{1.25}$Ga$_{0.75}$ from magnetization and first principle calculations.\cite{Enkovaara03} Here for Ni$_{1.75}$Mn$_{1.25}$Ga, the Ni magnetic moment (0.3$\mu_B$) is  parallel to MnII.  
The MnI and MnII moments for the anti-parallel (parallel) orientation are unequal:  -2.74 (1.9) and 3.23 (3.16) $\mu_B$, respectively. 
~ Thus, the magnetic moment of MnI is smaller than MnII. Smaller magnetic moment for MnI has also been obtained for Mn$_2$NiGa, and this has been assigned to stronger hybridization between the majority-spin Ni and MnII 3$d$ states in comparison to hybridization between Ni and MnI 3$d$ minority-spin states.\cite{Barman07} 

The difference in  total energy between the paramagnetic and ferromagnetic phases of Ni$_2$MnGa was equated to $k_B\,T_C$.\cite{Veliko99}  Following a similar approach,  the total energy difference between the ferro- and ferrimagnetic states (16 meV/atom) corresponds to 186\,K. As discussed earlier, $M(T)$  shows a decrease in magnetization at 120\,K  which is indicative of a magnetic transition. Since from theory, we find that the  MnI atoms have magnetic moment different from and anti-parallel to the MnII atom, we term the state below 120~K  to be ferrimagnetic. Here, anti-parallel alignment of unequal local Mn moments would exist for those MnII  atoms that have MnI as a nearest neighbor. 
~The estimate of a transition temperature of 186~K from theory can be considered to be in fair agreement with the experiment (120\,K), considering that  theory considers an ideal situation while the actual conditions may be more complicated. 
~For example, the MnI atoms would replace the Ni atoms at random positions, and absence of any superlattice peak in the XRD pattern (Fig.~3) indeed indicates that. This disorder effect in not considered in theory. Moreover, anti-site defects, possibility of canted alignment are not considered by theory. So, in reality, the lattice sites where anti-parallel alignment between MnI and MnII moments occurs would be random and the  moment of MnI could be less than what is calculated. 
In fact, this is indicated by the underestimation of the total moment by theory (3.1~$\mu_B$) 
~compared to the experimental value of 3.5~$\mu_B$ (Fig.~4b).




To explain the positive MR  shown in Figs.~1 and 2, we note that the application of magnetic field to a state with partial antiferromagnetic alignment of moments (MnI and MnII in this case) would induce spin fluctuations, thus increasing the spin disorder and hence resistivity that would result in positive MR.\cite{Bauer01,Molner81} In Eu$_{0.83}$Fe$_4$Sb$_{12}$, large positive MR has been assigned to a ferrimagnetic or canted magnetic phase.\cite{Bauer01} We also find that for Ni$_{1.75}$Mn$_{1.25}$Ga  positive MR increases linearly with field. 
~In many antiferromagnetic intermetallic alloys positive MR  has been observed to increase linearly with H.\cite{Sampath95} 
~For Eu$_{0.83}$Fe$_4$Sb$_{12}$, at low temperatures a H$^{2/3}$ variation was observed. Linear positive MR has been recently reported for Fe, Co and Ni thin films up to 60~T and has been explained by quantum electron-electron interaction theory.\cite{Greber07} To the best of our knowledge, no theoretical prediction exists about MR behavior for a ferrimagnet with  disordered antiferromagnetic alignment of a fraction of the local moments. To understand the linear MR variation in Ni$_{1.75}$Mn$_{1.25}$Ga, measurements with higher fields would constitute an interesting study. 

In the martensitic phase,  MR is negative and its magnitude increases  up to the saturation field (5 and 50~K data in Fig.~1b). But the behavior is clearly different from the austenitic phase: the slope of MR(H) does not change with temperature between 0 to 2~T (see the 5 and 50~K plots in Fig. 1b). In contrast, the slope changes between 300, 235 and 150~K data in the austenitic phase where $s$\,-\,$d$ scattering dominates.
This indicates that the origin of negative MR is different in the martensitic phase.  
Unlike in Ni$_2$MnGa,\cite{Biswas05} here the $T_C$ (=\,380~K) is much higher than $M_s$ (=\,76~K) and so the effect of $s-d$ scattering in the martensitic phase is not visible. 

One of the reasons for the increase in $\rho$ in the martensitic phase is the scattering of the Bloch wave functions at the twin boundaries (TB), which are known to increase the defect density and hence resistivity;\cite{Shatzkes73} and this has been reported earlier for FSMA's.\cite{Jin03,Sokhey03} The origin of negative MR in the martensitic phase that leads to positive to negative switching of MR while cooling (Fig.~1b and 2b) possibly arises from the decrease in electron scattering due to decrease in the density of twin boundaries and domain walls with the application of external magnetic field. These are oriented in dissimilar directions at zero field and would tend to form larger twin variants and domains as the saturation field is reached. This will have smaller resistivity compared to the twinned state with small sized twins and domains at $H$=\,0.\cite{Pan00,Biswas06} Negative MR due to domain wall scattering has been observed in ferromagnetic thin films.\cite{Gregg96,Ravelosona99,Gil05,Seemann07} The hysteresis normally observed in domain wall MR is related to the hysteresis of the $M-H$ curve. However, for Ni$_{1.75}$Mn$_{1.25}$Ga the $M-H$ curves   hardly exhibit any hysteresis (Fig.~5). For Ni$_{1.75}$Mn$_{1.25}$Ga, the observation that the increase in the negative MR magnitude occurs for fields less than equal to the saturation field (arrow, Fig.~1b), suggests that its origin is linked with the twin and domain rearrangement. 

Twin boundary motion occurs when twinning stress is small and MCA is high governed by the condition $K>\epsilon_0\,\sigma_{tw}$, where $K$ is the magnetic anisotropy energy density, $\sigma_{tw}$ is the twinning stress and $\epsilon_0$ is the maximal strain given by $(1-c/a)$.\cite{Sozinov02} For this specimen,  high MCA is expected because $M_s$ is much lower than $T_C$ and this is supported by magnetization data in Fig.~4a. In fact, the decrease in magnetization at $M_s$ gives rise to inverse magnetocaloric effect, and its magnetic field dependence has been explained by twin variant reorientation.\cite{Marcos02a}   From XRD, we find that the unit cell volume remains similar across the martensitic transition and the structure is modulated in the martensitic phase (Fig.~3). Modulated structures have lower twinning stress and hence are expected to exhibit twin boundary motion.\cite{Likhachev04}  
~These indicate that Ni$_{1.75}$Mn$_{1.25}$Ga would have small twinning stress and thus exhibit twin boundary motion. In fact, highest MFIS of 10\% has been reported in a Mn excess  specimen with composition of Ni$_{1.95}$Mn$_{1.19}$Ga$_{0.86}$ 
 ~that exhibits seven layer modulated structure and a low twinning stress of 2~MPa.\cite{Sozinov02}  MFIS has been reported to occur in polycrystals that are textured and with large grain size, in trained samples, and also in fine grained systems.\cite{Boon07,Sozinov04,Ullakko01,Jeong03,Pasquale00,Gaitzsch07} In our case,  the specimen has been annealed for more than a week at 1100~K, and that leads to the growth of  large grains (200-500 $\mu$m). On the other hand,  in the absence of field, the width of the twins is only a few microns.\cite{Pan00,Biswas06,Chopra00} Thus, within a grain the twins are ubiquitous and twin boundary rearrangement can occur due to external magnetic field. Coarse grained Ni-Mn-Ga is known to show larger MFIS, while  annealing of Ni-Mn-Ga ribbons is reported to increase the  MFIS by an order of magnitude.\cite{Soederberg01}  
 ~ 
~ Sozinov {\it et al.} obtained single variant state for polycrystalline Ni-Mn-Ga at 1~T.\cite{Sozinov04} For polycrystals, since the grains are oriented randomly, which lead to internal geometric constraints, for example, the motion of the twin boundaries would be suppressed by the grain boundaries, the macroscopic strain is  small. However, at the microscopic level within a grain the twin boundary motion is expected to occur in Ni$_{1.75}$Mn$_{1.25}$Ga and this is what is important in the present context to explain the negative MR behavior in the martensitic phase. The characteristic shape of the MR curve in the martensitic phase in polycrystalline Ni-Mn-Ga has been explained by twin variant rearrangement with magnetic field.\cite{Biswas05}

The difference in resistivity between the first and second FH cycles (Fig.~2) can be related to the extent of twin boundary rearrangement. Lower values of $\rho$$(T)$ in second FH cycle, which is recorded after cooling in presence of magnetic field across the martensitic transition from 300~K, indicate that the field would more effectively reorient the twins as soon as these are formed below $M_s$. On the other hand, for FH first cycle where $\rho$ is higher the magnetic field is switched on at 5~K. Thus,  although MCA increases with decreasing temperature,\cite{Albertini02} twinning stress may also increase limiting the extent of twin variant rearrangement forming larger twins. 
The variation of negative MR with temperature in the martensitic phase shown in Fig.~2b is significant because the heating and cooling data are very similar. This may be related to the microscopic details of the domain and twin variant reorientation with temperature and further studies are required to explain this.  

Another interesting observation in the MR of the martensitic phase is as follows: above the saturation field although MR is negative, a positive component is evident from its gradual increase with field (see Fig.~1b).  For example, at 50\,K MR increases from -0.2\% at 1\,T to -0.1\% at 8\,T. A weakly increasing MR is also observed for the 5\,K data, but its slope is less compared to 50\,K. This shows that at lower temperatures where thermal fluctuations decrease, higher field would be required to induce spin disorder in the ferrimagnetic state for causing similar increase in MR.  This, in turn, supports the argument that below 120\,K the magnetic state is ferrimagnetic in nature.
~Thus, there are two competing effects that govern MR in the martensitic phase with increasing field: a dominant negative trend due to the formation of larger size twin variants and domains and a weaker positive trend due to the ferrimagnetic nature of the magnetic state. While the first effect is present only up to the saturation field, the second effect becomes visible only above the saturation field. 

\section{Conclusion}
The switching of MR from negative to positive and back to negative values with decreasing temperature is observed in a Mn excess ferromagnetic shape memory alloy Ni$_{1.75}$Mn$_{1.25}$Ga. Positive  MR below 120\,K in the austenitic phase  is related to a ferrimagnetic state where the excess Mn atoms (MnI) at Ni site are antiferromagnetically oriented with the Mn atoms at Mn site (MnII). The existence of the ferrimagnetic state is shown by density functional theory, and experimental evidence is obtained from a decrease in magnetization and resistivity upturn  at 120~K. In the martensitic phase, negative  MR arises due to decrease in electron scattering related to reduction in the density of twin boundaries and domain walls with the application of external magnetic field. This effect is visible up to the saturation magnetic field. Above this, a weaker positive trend due to the ferrimagnetic nature of the magnetic state is visible. On the other hand, negative MR above 120\,K in the ferromagnetic austenitic phase is explained by the $s-d$ scattering model. The hysteresis in MR(T) is a manifestation of the first order nature of the martensitic phase transition.

\section{Acknowledgment}
We are grateful to  S. N. Kaul, A. K. Majumdar, E. V. Sampathkumaran, N. Lakshmi, A. Banerjee, R. Ranjan  and K. Maiti for useful discussions and support. P. Nordblad is thanked for his valuable support during the visit of PKM to his laboratory. P. Chaddah, K. Horn, A. K. Raychaudhuri, V. C. Sahni,  B. K. Sharma,  A. Gupta and S. M. Oak are thanked for constant encouragement.  Research grants from Ramanna Research Grant, Max Planck partner group project, Department of Science and Technology, India, and SIDA, Sweden are acknowledged. \\
\vskip 2cm

\noindent $^{*}$barman@csr.ernet.in

\newpage
\noindent{\bf Figure Captions:}\\

\noindent Fig.~1: Isothermal magnetoresistance (MR, $\Delta\rho_{m}$) of Ni$_{1.75}$Mn$_{1.25}$Ga, as a function of magnetic field at different temperatures (a) 300 to 150~K (b) 100 to 5~K. MR has been  measured in the cooling cycle. The arrow indicates the saturation magnetic field at 20~K. Solid lines are fit to the data.
\vskip .5cm
\noindent Fig.~2: (Color online) (a) Resistivity ($\rho$) as a function of temperature at zero and 5\,T field for Ni$_{1.75}$Mn$_{1.25}$Ga between 5 to 180\,K for two cooling and heating cycles. The data has been recorded in the following sequence: zero-field heating (ZFH); zero-field cooling (ZFC); the sample is then subjected to a magnetic field of 5T and field-cooled heating (FH, first cycle) is done up to RT; field-cooled cooling (FC) and once again FH (second cycle) was measured. $A_s$, $A_f$, $M_s$, and $M_f$ are the  austenitic and martensitic start and finish temperatures, respectively. (b) Magnetoresistance during cooling and heating calculated from the difference of  $\rho$$(T)$ at zero and 5\,T after interpolating to same temperature intervals. The difference has been taken between ZFC and FC in the cooling cycle and ZFH and FH second cycle while heating.
\vskip .5cm
\noindent Fig.~3: X-ray diffraction pattern  of Ni$_{1.75}$Mn$_{1.25}$Ga as a function of temperature while cooling. The calculated profiles obtained by Le Bail fitting (solid line through the experimental data points) are shown for 150 and 40~K data. $A$ and $M$ indicate the peaks related to the austenitic and martensitic phases, respectively.
\vskip .5cm
\noindent Fig.~4: (a) Temperature dependence of magnetization in (a) a small applied field of 0.01~T 
\vskip .5cm
\noindent Fig.~5: Isothermal $M-H$ loops at different temperatures.


\begin{thebibliography}{}
\bibitem{Baibich88}M. N. Baibich,  J. M. Broto, A. Fert, F. Nguyen Van Dau, F. Petroff, P.Eitenne, G. Greuzet, A. Friederich, and J. Chazelas, Phys. Rev. Lett. {\bf61}, 2472 (1988); 
~S. Jin, T. H. Tiefel, M. McCormack, R. A. Fastnacht, R. Ramesh, and L. H. Chen, 
  Science {\bf 264}, 413 (1994).
\bibitem{Kainuma06} R. Kainuma, Y. Imano, W. Ito, Y. Sutou, H. Morito, S. Okamoto, O. Kitakami, K. Oikawa, A. Fujita, T. Kanomata and K. Ishida, Nature  {\bf439},  957 (2006).
\bibitem{Krenke05} T. Krenke, E. Duman, M. Acet, E. F. Wassermann, X. Moya, L. Ma$\tilde{n}$osa and A. Planes, Nature Mat. {\bf4}, 450 (2005).
\bibitem{Takeuchi03} I. Takeuchi, O. O. Famodu, J. C. Read, M. A. Aronova, K. -S. Chang, C. Craciunescu, S. E.  Lofland, M. Wuttig, F.C Wellstood, L. Knauss and A. Orozco, Nature Materials {\bf 2}, 180 (2003).  
\bibitem{Planes97} A. Planes, E. Obrad$\acute{o}$, A. Gonz$\grave{a}$lez-Comas, and L. Ma$\tilde{n}$osa, 
~Phys. Rev. Lett. {\bf 79}, 3926 (1997).
\bibitem{Boon07} Y. Boonyongmaneerat, M. Chmielus, D. C. Dunand, and P. M$\ddot{u}$llner, Phys. Rev. Lett. {\bf 99}, 247201 (2007).
\bibitem{Chernenko04} V. A. Chernenko, V. A. L$'$vov, P. M\"{u}llner, G. Kostorz and T. Takagi, Phys. Rev. B {\bf 69}, 134410 (2004).
\bibitem{Barman05} S. R. Barman, S. Banik, and A. Chakrabarti, Phys. Rev. B {\bf 72}, 184410 (2005).
\bibitem{Chakrabarti05} A. Chakrabarti,  C. Biswas, S. Banik, R. S. Dhaka, A. K. Shukla, and S. R. Barman, 
~Phys. Rev. B {\bf72}, 073103 (2005).
\bibitem{Barman07} S. R. Barman,  S. Banik, A. K. Shukla, C. Kamal, and A. Chakrabarti,~Europhys. Lett. {\bf80}, 57002 (2007); S. R. Barman and A. Chakrabarti, Phys. Rev. B {\bf 77}, 176401 (2008).
\bibitem{Ahuja07}B. L. Ahuja, B. K. Sharma, S. Mathur, N. L. Heda, M. Itou, A. Andrejczuk, Y. Sakurai, Aparna Chakrabarti,
S. Banik, A. M. Awasthi, and S. R. Barman, Phys. Rev. B {\bf75}, 134403(2007).
\bibitem{Banik06} S. Banik, A. Chakrabarti, U. Kumar, P. K. Mukhopadhyay, A. M. Awasthi, R. Ranjan, J. Schneider, B. L. Ahuja, and S. R. Barman, Phys. Rev. B {\bf 74}, 085110 (2006).
\bibitem{Murray00} S. J. Murray, M. Marioni, S. M. Allen, R. C. O'Handley and T. A. Lograsso, 
~Appl. Phys. Lett. {\bf77}, 886 (2000).
\bibitem{Sozinov02} A. Sozinov, A. A. Likhachev, N. Lanska and K. Ullakko, Appl. Phys. Lett. {\bf80}, 1746 (2002).
\bibitem{Albertini01} F. Albertini, L. Morellon, P. A. Algarabel, M. R. Ibarra, L. Pareti, Z. Arnold and G. Calestani, 
~J. Appl. Phys. 89, 5614 (2001).
\bibitem{Ullakko01} K. Ullakko, Y. Ezer, A. Sozinov, G. Kimmel, P. Yakovenko, Y. K. Lindroos, Scripta Materialia {\bf44}, 475 (2001)
\bibitem{Sullivan04} M. R. Sullivan and H. D. Chopra, Phys. Rev. B {\bf 70}, 094427 (2004).
\bibitem{Sozinov04} A. Sozinov, A. A. Likhachev, N. Lanska, O. S$\ddot{o}$derberg, K. Ullakko, V.K. Lindroos,  Mat. Sci. Eng. A {\bf 378} 399 (2004).
\bibitem{Pan00} Q. Pan and R. D. James, J. Appl. Phys. {\bf 87}, 4702 (2000).
\bibitem{Marcos02} J. Marcos, A. Planes, L. Ma$\tilde{n}$osa, A. Labarta, B. J. Hattink, 
~Phys. Rev. B {\bf 66}, 054428 (2002).
\bibitem{Biswas05} C. Biswas, R. Rawat, and S. R. Barman, Appl. Phys. Lett. {\bf 86}, 202508 (2005). 
\bibitem{Lund02}M. S. Lund, J. W. Dong, J. Lu, X. Y. Dong, C. J. Palmstrom, and C. Leighton, 
~Appl. Phys. Lett. {\bf80}, 4798 (2002).
\bibitem{Golub04} V. O. Golub, A. Ya. Vovk, L. Malkinski,  C. J. O'Connor, Z. Wang and J. Tang, J. Appl. Phys. {\bf96}, 3865 (2004) 
\bibitem{Yu06} S. Y. Yu, Z. H. Liu, G. D. Liu, J. L. Chen, Z. X. Cao, G. H. Wu, B. Zhang and X. X. Zhang, Appl. Phys. Lett. {\bf89} 89, 162503 (2006).
\bibitem{Sharma06} V. K. Sharma, M. K. Chattopadhyay, K. H. B. Shaeb, Anil Chouhan, and S. B. Roy, Appl. Phys. Lett. {\bf89}, 222509 {2006}.
\bibitem{Liu05} G. D. Liu, J. L. Chen, Z. H. Liu, X. F. Dai, G. H. Wua, B. Zhang and X. X. Zhang, 
~Appl. Phys. Lett.  {\bf87}, 262504  (2005); Appl. Phys. Lett.  {\bf82}, 424  (2003).
\bibitem{Enkovaara03} J. Enkovaara, O. Heczko, A. Ayuela, and R. M. Nieminen, 
~Phys. Rev. B {\bf67}, 212405 (2003).
\bibitem{Banik07a} S. Banik, R. Ranjan, A. Chakrabarti, S. Bhardwaj, N. P. Lalla, A. M. Awasthi, V. Sathe, D. M. Phase, P. K. Mukhopadhyay, D. Pandey, and S. R. Barman, 
~Phys. Rev. B {\bf75}, 104107 (2007).
\bibitem{Rajeev} R. Rawat and I. Das, J. Phys. Condens. Matter {\bf 13}, L1 (2001); A. Venimadhav, M. S. Hedge, R. Rawat, I. Das, P. L. Paulose, and E. V. Sampathkumaran, Phys. Rev. B {\bf63}, 214404 (2001).
\bibitem{Wien97} P. Blaha, K.  Schwartz, J. Luitz,  WIEN97, A Full Potential Linearized Augmented Plane Wave Package for Calculating Crystal Properties  
~ (Karlheinz Schwarz, Tech. Universit\"{a}t, Austria), ISBN 3-95010310-4.
\bibitem{Perdew96} J. P. Perdew and K. Burke, Phys. Rev. Lett. {\bf 77}, 3865 (1996). 
\bibitem{Kataoka01} M. Kataoka, Phys. Rev. B {\bf63}, 134435 (2001).
\bibitem{Hurd81} C. M. Hurd, I. Shiozaki and S. P. McAlister, Phys. Rev. B. {\bf 26}, 701 (1982). 
~
~
\bibitem{Lebail88} A. Le Bail {\it et al.},  Mat. Res. Bull. {\bf23},   447 (1988).
\bibitem{Brown02}  P. J. Brown, J. Crangle, T. Kanomata, M. Matsumoto, K-U. Neumann, B. Ouladdiaf and K. R. A. Ziebeck, J. Phys.: Condens. Matter {\bf14}, 10159 (2002); J. Pons, R. Santamarta, V. A. Chernenko, E. Cesari, J. Appl. Phys. {\bf 97}, 083516 (2005); R. Ranjan, S. Banik, U. Kumar, P.K. Mukhopadhyay, S.R. Barman, D. Pandey, Phys. Rev. B {\bf74}, 224443 (2006); L. Righi, F. Albertini, G. Calestani, L. Rareti, A. Paoluzi, C. Ritter, P. A. Algarabel, L. Morellon, and M. R. Ibarra, J. Solid State Chem. {\bf179}, 3525 (2006)
\bibitem{Bhattacharya03} K. Bhattacharya, {\it Microstructure of Martensite. Why it Forms and How it Gives Rise to the Shape-Memory Effect} (Oxford University Press, Oxford, 2003).
\bibitem{Albertini02} F. Albertini, L. Pareti, and A. Paoluzi, L. Morellon, P. A. Algarabel, M. R. Ibarra, and L. Righi, Appl. Phys. Lett., {\bf81}, 4032 (2002).
\bibitem{Webster84} P. J. Webster, K. R. A. Ziebeck, S. L. Town, M. S. Peak, Philos. Mag. {\bf49}, 295 (1984).
\bibitem{Hobbs03} D. Hobbs, J. Hafner, D. Spisak, Phys. Rev. B {\bf 68}  014407 (2003).
\bibitem{Veliko99} O.I. Velikokhatny$\check{i}$ and I. I. Nuamov, Phys. Solid State {\bf41}, 617(1999).
\bibitem{Bauer01} E. Bauer, St. Berger, A. Galatanu, M. Galli, H. Michor, G. Hilscher,  Ch. Paul, B. Ni, M. M. Abd-Elmeguid, V. H. Tran, A. Grytsiv, and P. Rogl, Phys. Rev. B.  {\bf 63},224414 (2001). 
\bibitem{Molner81} S. von Molner, R. J. Gambino, and J. M. D. Ceoy, J. Appl. Phys {\bf 52}, 2193 (1981); A. Fert and R. Asomoza, J. Appl. Phys. {\bf 50}, 1886 (1979).
\bibitem{Sampath95} E. V. Sampathkumaran and I. Das, Phys. Rev. B. {\bf 51}, 8631(1995); C. Majumdar, A. K. Nigam, R. Nagarajan, C. Godart, L. C. Gupta, B. D. Padalia, G. Chandra and R. Vijayaraghavan, Appl. Phys. Lett. {\bf 68}, 3647 (1996); A. K. Nigam, S. B. Roy and P. Chaddah, Phys. Rev. B. {\bf 60}, 3002 (1999); S. Radha, S. B. Roy and A. K. Nigam, J. Appl. Phys {\bf 80}, 6803 (2000).
\bibitem{Greber07} A. Gerber, I. Kishon, I. Ya. Korenblit, O. Riss, A. Segal, M. Karpovski, B. Raquet, Phys. Rev. Lett. {\bf99}, 027201 (2007).
\bibitem{Shatzkes73} M. Shatzkes, P. Chaudhari, A. A. Levi, A. F. Mayadas, Phys. Rev. B {\bf7}, 5058 (1973). 
\bibitem{Jin03} X. Jin, D. Bono, C. Henry, J. Feuchtwanger, S. M. Allen, and R. C. O'Handley, Philos. Mag. {\bf83}, 3193 (2003).
\bibitem{Sokhey03} K. S. Sokhey, M. Manekar, M. K. Chattopadhyay, R. Kaul, S. B. Roy, and P. Chaddah, J. Phys. D {\bf36}, 1366 (2003).
\bibitem{Biswas06}C. Biswas, S. Banik, A. K. Shukla, R. S. Dhaka, V. Ganesan and S. R. Barman, Surf. Sci. {\bf 600}, 3749(2006).
\bibitem{Gregg96}J. F. Gregg, W. Allen, K. Ounadjela, M. Viret, M. Hehn, S. M. Thompson, and J. M. D. Coey, Phys. Rev. Lett. {\bf 77}, 1580 (1996).
\bibitem{Ravelosona99} D. Ravelosona, A. Cebollada, and F. Briones, C. Diaz-Paniagua, M. A. Hidalgo, and F. Batallan, Phys. Rev. B {\bf 59}, 4322 (1999).
\bibitem{Gil05} W. Gil, D. G$\ddot{o}$rlitz, M. Horisberger, and J. K$\ddot{o}$tzler, Phys. Rev. B {\bf 72}, 134401 (2005).
\bibitem{Seemann07} K. M. Seemann, V. Baltz, M. MacKenzie, J. N. Chapman, B. J. Hickey, and C. H. Marrows, Phys. Rev. B {\bf 76}, 174435 (2007).
\bibitem{Marcos02a} J. Marcos, A. Planes, and L. Ma$\tilde{n}$osa, F. Casanova, X. Batlle, and A. Labarta, B. Mart$\acute{i}$nez, Phys. Rev. B {\bf66}, 224413 (2002). 
\bibitem{Likhachev04} A. A. Likhachev, A. Sozinov, and  K. Ullakko, Mater. Sci. Eng. A {\bf378}, 513 (2004).
\bibitem{Jeong03}S. Jeong, K. Inoue, S. Inoue, K. Koterazawa, M. Taya and K. Inoue, Mat. and Eng. A {\bf359}, 253 (2003). 
\bibitem{Pasquale00}M. Pasquale, C. Sasso, S. Besseghini, F. Passaretti, E. Villa, and A. Sciacca, IEEE Trans. Mag. {\bf36}, 3263 (2000).
\bibitem{Gaitzsch07}U. Gaitzsch, M. Potschke, S. Roth, B. Rellinghaus and L. Schultz, Scripta Mater. {\bf57}, 493 (2007).
\bibitem{Chopra00}H. D. Chopra and C. Ji, Phys. Rev. B {\bf 61}, R14 913 (2000)
\bibitem{Soederberg01} O. S$\ddot{o}$derberg, Y. Ge, N. Glavatska, O. Heczko, K. Ullakko, and V. K. Lindroos
J. Phys IV (France) {\bf 11}, 287 (2001); S. Guo, Y. Zhang, B. Quan, J. Li, and X. Wang, Mat. Sci. Forum, {\bf 475-479}, 2009 (2005).

\end{thebibliography}
\end{document}